\documentclass[useAMS,usenatbib]{mn2e}

\usepackage{longtable}
\usepackage{pdflscape}
\usepackage{graphicx}
\usepackage{pdflscape}
\usepackage{lscape,graphicx,pifont}
\usepackage{color}

\newcommand{\Msun}{\mbox{\,M$_\odot$}}

\newcommand{\Lsun}{\mbox{\,L$_\odot$}}
\newcommand{\vunit}{\mbox{\,km\,s$^{-1}$}}
\newcommand{\mic}{\mbox{$\,\mu$m}}
\newcommand{\pion}[2]{{#1}\,{\sc {#2}}}
\newcommand{\fion}[2]{[{#1}\,{\sc {#2}}]}
\newcommand{\nucl}[2]{\mbox{$^{#1}${#2}}}

\newcommand{\gtsimeq}{\raisebox{-0.6ex}{$\,\stackrel
        {\raisebox{-.2ex}{$\textstyle >$}}{\sim}\,$}}

\newcommand{\aabo}{\mbox{AT2026aabo}}
\newcommand{\resp}[1]{\textcolor{black}{#1}}

\title[AT2026aabo: a dusty nova in M31]{AT2026aabo: a luminous dusty  nova in M31}

\author[A. Evans et al.]{A. Evans$^{1}$\thanks{E-mail: a.evans@keele.ac.uk},
D. P. K. Banerjee$^2$,
T. R. Geballe$^3$,
C. E. Woodward$^4$,  % \newauthor
S. Starrfield$^5$  \\
\mbox{} \\
$^{1}$Astrophysics Research Centre, Lennard Jones Laboratory, Keele
University, Keele, Staffordshire. ST5 5BG, UK\\
$^{2}$Physical Research Laboratory, Navrangpura,  Ahmedabad, Gujarat 
380009, India \\ % DPKB
$^3$Gemini Observatory/NSF's NOIRLab, 670 N. A`ohoku Place, Hilo, HI 96720 USA \\
$^4$Minnesota Institute for Astrophysics, School of Physics \& Astronomy,
116 Church Street SE, University of Minnesota, \\
Minneapolis, MN 55455, USA\\ %RDG, CEW
$^5$School of Earth and Space Exploration, Arizona State University, 
Box 871404, Tempe, AZ 85287-1404, USA\\ % SS
}
\begin{document}

\date{Version of \today}

\pagerange{\pageref{firstpage}--\pageref{lastpage}} \pubyear{2021}

\maketitle

\label{firstpage}

\begin{abstract}
We present near-infrared spectroscopy of the M31 transient
\aabo, obtained 36 and 59 days post eruption.
This is the first, and earliest, multi-epoch study of
a dust-forming event in an M31 nova. With the exception
of the \pion{O}{i}  line at 1.1289\mic\
and a weak Pa$\beta$ line on day~36,
the spectra are featureless, with evidence for 
the presence only of cooling dust. The mass of dust
is $\sim10^{-7}$\Msun\ \resp{for either amorphous
carbon or silicate dust}, while
the dust temperatures (1100~K on day~36, 750~K on
day 59) imply a bolometric luminosity for the 
stellar remnant of $\sim5-8\times10^5$\Lsun.
These values lie in the range of Galactic dusty
novae and are significantly lower than those
of luminous red novae. We conclude that \aabo\ is a luminous 
classical nova.
%% 124 words
\end{abstract}


\section{Introduction}
\label{intro}
\begin{figure}
\centering
 \includegraphics[width=0.45\textwidth]{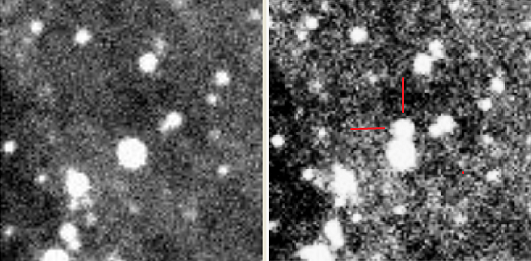}
  \caption{Left: the field of \aabo\ on MJD 61266, 
 pre-eruption. Right, \aabo\ is identified by the red markers 
 in an image taken on 61286.98 (2026 September 3.98) when 
 the nova was at maximum (14.97~mag in the ATLAS ``orange''
 band). Both images are from ATLAS. The field of view is
 $3'\times3'$.  North is up, east is left.
  \label{field}}
\end{figure}

The transient \aabo\ was discovered on
2026 August 29.395 at an AB $g$ magnitude of 18.58
\citep{salgundi26}; it was not detected on
2026 August 25.474 to a limiting AB $r$ magnitude of 20.83.
Photometry was reported by \cite{perez26}, who noted a
rapid brightening of the object and that
it is 1461\farcs8 east and 1022\farcs0
north of the center of M31. \citeauthor{perez26} describe
\aabo\ as a ``a bright nova candidate in M31''.
The field is shown in Fig.~\ref{field}.

\begin{figure}
\centering
 \includegraphics[width=0.45\textwidth]{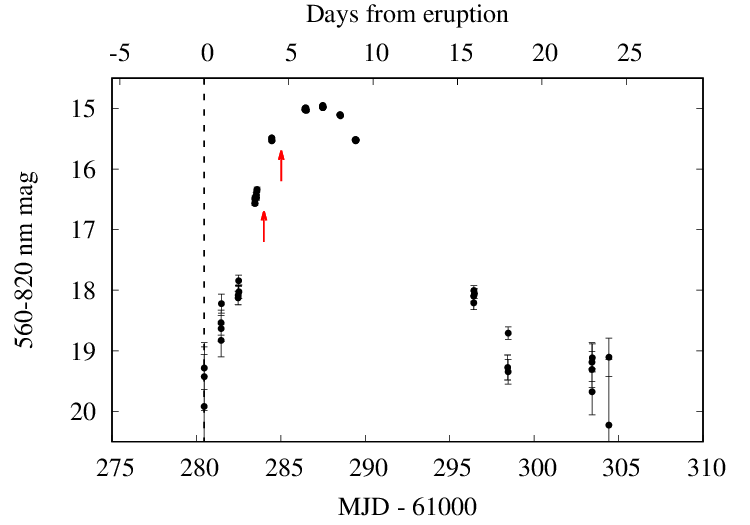}
 \caption{The light curve of \aabo\ in the ATLAS
 ``orange'' band; data from \protect\cite{perez26}
 and the ATLAS website. 
 The red arrows denote times of optical spectroscopy
 \citep{chamoli26,katusiime26} identifying the object
 as a nova. The vertical dotted line is the assumed
 origin of time. See text for details.  \label{LC}}
\end{figure}

The light curve in the ``orange'' Asteroid Terrestrial-impact
Last Alert System \citep[ATLAS;][]{tonry18} filter band,
which covers the wavelength range 560--830~nm,
is shown in Fig.~\ref{LC}.
The object had brightened to $15.05\pm0.01$ on
2026 September 2.699 \citep{chamoli26}.
The rapid rise, by $\sim4$~magnitudes in $\sim4$~d,
is typical of the pre-maximum rise of a nova
\citep[see][for pre-maximum light curves 
of novae]{hounsell10,mcquillin12}. 
\resp{The rise to maximum, magnitude at peak, and
the rate of decline, are similar to those of the 
luminous M31 nova M31N 2007-11d \citep{shafter09}.}

Fitting a parabola to the peak of the light curve
gives maximum light at MJD 61287. A linear fit to
the post-maximum light curves gives $t_2$
(the time to decline by 2~magnitudes from
maximum) to be 4.64~d, and $t_3=6.96$~d.
For our present purpose, we take the time of
eruption to be MJD~61280.461466, the time of the first
photometric data point in Fig.~\ref{LC}.

\begin{figure*}
 \includegraphics[width=0.5\textwidth]{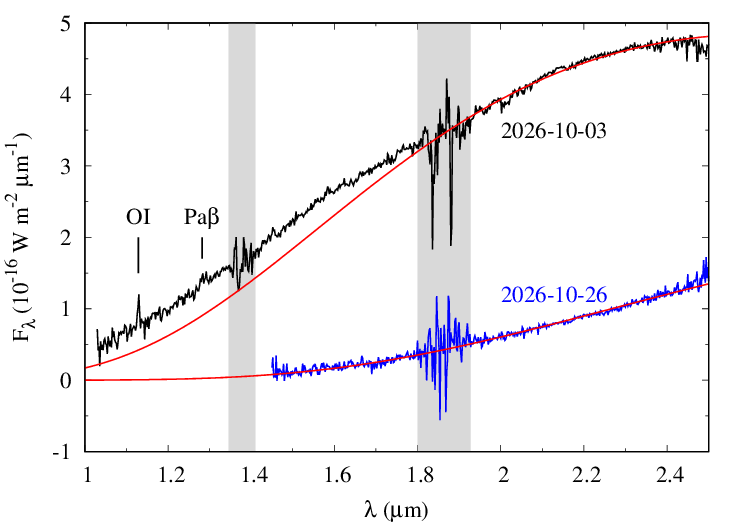}
 \caption{Observed (i.e. not dereddened) Gemini North spectra of \aabo.
 Grey bars indicate wavelength ranges where the 
 atmospheric transmission is poor.
 The red curves are black body fits to the respective continua,
 and reddened by $E(B-V)=0.1$. See text for details.
 \label{GN}}
\end{figure*}

The identifcation of AT2026aabo as a ``normal''
nova seemed to be confirmed
by optical spectroscopy obtained on days 3.5 and 4.5 \citep{chamoli26}.
Their spectra showed a velocity shift of $\sim300$\vunit,
almost identical to the heliocentric radial velocity of
M31 \citep[][]{courteau99}. This is consistent with \aabo\ being
located in M31. The spectra showed prominent 
blueshifted \pion{Fe}{ii} lines, and P-Cygni profiles
in the Balmer lines, typically seen in novae around 
photometric maximum. A somewhat later
(day~7.5) optical spectrum \citep{katusiime26}
also showed prominent \pion{Fe}{ii} and \pion{O}{i} lines,
together with broad \pion{H}{i} emission lines. 
The H$\alpha$ absorption extends blueward to 2000\vunit. 
The \pion{He}{i} lines in the spectrum 
reported\footnote{Available at https://www.wis-tns.org/object/2026aabo.}
by \cite{katusiime26} also contains some evidence of
P~Cygni profiles.
These observations indicate that during that time period,
\aabo\ was in the \pion{Fe}{ii} phase. 
\resp{In this respect, \aabo\ resembled the luminous
M31 novae M31N 2016-08d \citep{williams16} and the possible nova
PNV J00432114+4124597 \citep{darnley15}.}

Taking the distance of M31 to be 770~kpc, and
the \resp{foreground (Galactic)} reddening of M31
(and hence of \aabo) to
be $E(B-V)=0.1$ \citep{darnley17},
the absolute $g$ magnitude at maximum would have
been $M_g = -9.22$, making \aabo\ one of the brightest
novae in M31 \citep{clark24}.
The bolometric correction for a nova at maximum
is not known for the $g$ band, but it is unlikely to
differ significantly from the value $-0.15$ for the 
$V$ band \citep{vanden87,selvelli19}. We therefore
assume $M_{\rm bol}
=-9.37$, corresponding to $L_{\rm bol}=4.4\times10^5$\Lsun\
(cf. the Eddington luminosity, $4.72\times10^4$\Lsun, for a
1.44\Msun\ white dwarf).
\resp{The absolute magnitude of \aabo\ at maximum is
similar to those of the M31 novae M31N 2005-01a 
and M31N 2009-10b \citep{shafter11b}.}

However, \cite{mujortova26}, on the basis of an optical
spectrum obtained on September 3.67 UT, concluded that the 
properties of \aabo\ were also consistent with those
expected for a ``luminous red nova'' 
\citep[LRN; see][hereafter KB26, for a review]{kaminski26}
during its early evolution. We discuss this interpretation
below.

\cite{karambelkar26} reported that \aabo\ had brightened 
significantly in the near-infrared (NIR), 
from $H_s = 17.39\pm0.03$ (AB)
on 2026 September 11 UT to $H_s = 16.71\pm0.02$~mag (AB) on 
2026 September 16 UT (day~18). At the same time, the object
reddened substantially, from $(J-H_s)\sim -0.1$~mag
to $(J-H_s)\sim+0.5$~mag. Contemporaneous observations from
ATLAS showed that the source continued to fade in the optical
bands during the same period, reddening to $(c-H_s)\sim2.5$~mag 
\citep[where $c$ is the ATLAS ``cyan'' magnitude; see][]{tonry18} 
on MJD 61298.

Here we present NIR spectroscopy of \aabo\ that
sheds additional light on its nature.

\section{Observations and data reduction}

NIR spectra of \aabo\ covering 0.8--2.5\mic\ were
obtained at the Frederick C. Gillett Gemini North
telescope on the nights of UT 2026 October~3
(MJD 61316, $t=36$~d) and October~26
(MJD 61339, $t=59$~d), using the Gemini 
Near-Infrared Spectrograph \citep[GNIRS;][]{elias06} in 
cross-dispersed mode\footnote{Attempts to obtain NIR
spectra at earlier times were thwarted by poor weather.}.
On both occasions the instrument was
configured with its 32~$\ell$~mm$^{-1}$ grating and its slit of
dimensions $0\farcs675\times7''$.  This produced a nominal
spectral resolving power ($R = \lambda/\delta\lambda$) of
800 across the entire wavelength range. 

Once \aabo\ was centred in the slit, spectra were obtained in the
standard ABBA nod-along-slit mode with nod angles
of $\pm1\farcs5$.  On each night individual exposures were 200~s,
and the total exposure time was 3200~s. Nearby A0V stars, 
observed at comparable airmasses as those for \aabo, were
employed as flux and telluric standards. The spectrum of
an argon arc  lamp served as the wavelength calibrator;
the accuracy of the calibration is better than 0.0005\mic.

On October 3, the $H$ magnitude of \aabo, as estimated from
the reduced spectrum, was $\sim16.6$ and it was possible
to acquire by direct imaging. On October 26, however, it had
faded by nearly three magnitudes and could not be identified
in the crowded field by direct imaging. Therefore, the
object was acquired using a blind offset from a Gaia
star only $18''$ distant from the science target. 

Data reduction, incorporating both IRAF \citep{tody93}
and Figaro \citep{currie14} routines, proceeding as
described in \cite{evans22}. On October 3, the spectrum
of the science target shortward of 1.0\mic\ was not detected.
For the October 26 observations, neither the continuum nor line emission was detected shortward of 1.5\mic.  

On the GNIRS array, when viewing a spectrum produced by the
32~$\ell$~mm$^{-1}$ grating, the pixel spacing in wavelength
is $\lambda/1800$, ranging linearly from $\sim0.0006$\mic\
at 1.0\mic\ to $\sim0.0014$\mic\ at 2.5\mic. For the 0\farcs675
slit, the wavelength
resolution ($\lambda/R$) ranges linearly from
0.00125\mic\ at 1.0\mic\ to 0.0031\mic\ at 2.5\mic. 
Because of the low signal-to-noise ratios of the final spectra,
Fig.~\ref{GN} covers only the NIR wavelengths where
signal from \aabo\ was detected ($>3\sigma$). 
The spectra are plotted in bins of width 0.002\mic.
The selected bin width undersamples the spectral resolution at
wavelengths $<1.6$\mic. This is unimportant, as the
only detected lines in the spectrum are much wider
than 0.002\mic. 

\section{Results}

The two GNIRS spectra are shown in Fig.~\ref{GN}. One
characteristic is immediately evident: with the
exception of \pion{O}{i} $^3$P$-^3$D$^o$ at 1.1289\mic\
and a weak Pa$\beta$, {\it both spectra are dominated by
dust emission and are in general devoid of emission lines}
\citep[see Table~1 of][for a list of NIR emission lines
commonly seen in erupting novae]{banerjee12}.
In particular, the \pion{He}{i} triplet at 1.0833\mic,  
which can be fairly strong in the spectra of erupting
novae, is conspicuous by its absence. 

We have fitted black body functions to the dereddened
novae spectra obtained at day 36 and day 59.  The fit to
the later epoch, over the entire dataset, gives a black body
temperature of 750~K.
The fit to the earlier epoch is rather better if the fit
is confined to the longest wavelengths (1.9--2.5\mic); 
the black body temperature in this case is 1100~K. 
The fits (reddened by $E(B-V)=0.1$ to 
match the observed data), are shown in Fig.~\ref{GN}.

The dust temperature $T_{\rm d}$ can be used to estimate
the bolometric luminosity $L_{\rm bol}$ of the heating
source, using
\begin{equation}
 T_{\rm d} = \left ( 
 \frac{L_{\rm bol} \langle{Q}_{\rm abs}(T_*)\rangle}
 {16\pi{r}^2\sigma\langle{Q}_{\rm abs}(T_{\rm d})\rangle} \right )^{1/4} \:\:. \label{Td}
\end{equation}
Here the $\langle{Q}_{\rm abs}(T)\rangle$ are the Planck
mean absorption efficiencies for the grain material at the 
stellar ($T_*$) and dust temperatures, respectively,
$r$ is the distance of the dust from the star at time $t$,
and $\sigma$ is the Stefan-Boltzmann constant.
Assuming that the dust is amorphous carbon, we use 
$\langle{Q}_{\rm abs}(T_{\rm d})\rangle = 58.16aT_{\rm d}^{0.754}$,
$a$ being the grain radius in cm \citep{evans17}, and 
$\langle{Q}_{\rm abs}(T_*)\rangle=1$.
We take the outflow velocity to be 2000\vunit\ 
\citep{katusiime26}, and $a=1$\mic\ 
\citep[e.g.,][]{evans17}. From Eq.~(\ref{Td})
\begin{eqnarray}
 L_{\rm bol} & \simeq & 4.95\times10^{14} 
 \left [ \frac{V}{2000\vunit} \right ]^2
 \left [ \frac{t}{\mbox{d}} \right ]^2 
 \left [ \frac{a}{1\mic}\right ]
 T_{\rm d}^{4.754}   \nonumber \\
 && \label{Lbol}
\end{eqnarray}
yielding a bolometic luminosity of 
$L_{\rm bol}\simeq1.85\times10^{32}$~W 
($4.82\times10^5$\Lsun,
$M_{\rm bol}\simeq-9.47$)
on day~36, and $L_{\rm bol}\simeq8.02\times10^{31}$~W
($2.10\times10^5$\Lsun, $M_{\rm bol}\simeq-8.56$) on day~59.
Given the approximations, these values
are consistent with the location of \aabo\ in M31
(see Section~\ref{intro}),
and place \aabo\ towards the upper
end of the bolometric luminosities of novae in M31
\citep{kasliwal11,clark24}. 
\resp{The derived values are not sensitive to the
assumed nature of the dust. Had
we assumed silicate dust, the luminosity
on day 36 (day 59) would be $1.83\times10^5$\Lsun\
($1.06\times10^5$\Lsun), where we have used 
temperature-dependent Planck means
for 1\mic\ silicate grains from \cite{laor93}.}

Using the formalism outlined in \cite{evans17},
and assuming amorphous dust with density 1.5~gm~cm$^{-3}$,
the dust mass is $3.65\times10^{-8}$\Msun\ for the 
first epoch, and $1.52\times10^{-7}$\Msun\ for the
second. 
\resp{Had we assumed silicate grains, the dust mass would be
$2.54\times10^{-7}$\Msun\ (day 36) and
$1.17\times10^{-6}$\Msun\ (day 59); note that the emission
efficiency is lower for silicates than it is for amorphous
carbon, so more silicate dust is required to produce the
observed flux.}
These values are in the range of dust masses
in the ejecta of novae \citep[see Table~1 in][]{evans25},
and very much lower than the masses of dust typically formed
in LRNe events ($\sim10^{-5}-10^{-3}$\Msun; see Table~1 in KB26).

There is no evidence in our NIR spectra for the presence
of first overtone CO (\nucl{12}{C}\nucl{16}{O} 2--0 band
head at 2.2935\mic) either in emission or absorption; see
Fig.~\ref{GN}. However even if CO were present, it is 
unlikely to be detected in our data. The vibrational levels
$\upsilon=2, 3, 4$, etc., from which the 2--0, 3--1, 4--2,
etc., band head emission arises, would not be excited unless
the vibrational temperature were at least 2000~K. For the
case of absorption, for the 0--2 band head to be detected, the
rotational levels that would need to be excited in the $v=0$
state are about $J\simeq50$, and the gas temperature would
need to be $\gtsimeq2000$~K for them to be significantly 
populated.

\section{Discussion and conclusions}

The observational evidence (in the forms of limited light curve
and spectroscopy) available to deduce the nature of \aabo\ is
extremely sparse. Its location in M31 seems uncontroversial,
on the basis of its heliocentric radial velocity, giving
an absolute magnitude of $\sim-9.4$. The dust emission
suggests a similar (bolometric) luminosity.

Our data rule out the possibility that \aabo\
is a LRN, as suggested by \cite{mujortova26}, for the 
following reasons. First, the bolometric luminosity
is towards the lower end of, or even below, the range
shown by LRNe (see
Figure~1 of KB26). Second, the deduced dust masses
($\sim10^{-7}$\Msun) are considerably lower than those
typically seen in LRNe (see Table~1 of KB26). Third,
the timescale for the event is $\sim20$~d,
whereas the timescale for LRNe is generally in excess of
$\sim100$~d \citep[][KB26]{kasliwal12}.
Finally, the light curve (Fig.~\ref{LC}) is incompatible
with a LRN. \resp{Indeed, comparison of the shape,
and time scale, of the light curve in Fig.~\ref{LC} and that 
of the M31 LRV AT2025abao \citep[taken as typical of the type;
see Figure~2 of][]{reguitti26}
shows clearly that \aabo\ cannot be a LRV.}

The early optical spectra \citep{chamoli26,katusiime26} 
seemed consistent with those of novae close to maximum,
with hydrogen recombination lines, \pion{Fe}{ii} lines, and
\pion{He}{i} lines, together with
evidence for material outflowing with velocities up
to $\sim2000$\vunit, rather higher than is typical
for Galactic dust-forming novae. They
that an eruption was under way, and that the ejected gas
was being excited by an ultraviolet\footnote{Unfortunately, 
an approved Target-of-Opportunity
proposal to observe \aabo\ with the Neil Gehrels Swift 
observatory was scheduled just as the Swift mission
came to an end.} (UV) source capable of ionising H.

By the time of our NIR observations, 
however, NIR emission lines (with the exception of
the \pion{O}{i} and Pa$\beta$ lines) had disappeared. 
On day 36, the \pion{O}{i} line was noticeably
stronger than Pa$\beta$, as is usually the case 
during the first $\sim15$~d of the evolution of
\pion{Fe}{ii} type novae \citep{banerjee12},
and also at later times \citep*{das09}.
It is not unusual for emission
lines to fade as the dust condenses in dust-forming
novae. This is well illustrated in the case of V1280~Sco
\citep[see Figure~2 of][]{das08},
in which dust started forming around day~24;
before this time the NIR spectrum was dominated
by line emission, but weakened as dust formed.
Similar behaviour was seen in V2615~Oph,
which formed dust around day~85, before which
emission lines dominated the NIR spectrum,
but thereafer the lines faded \citep{das09}.

To date, there have been only two dust detections in M31
novae, both of which were of the \pion{Fe}{ii} class. 
These detections were from single epoch ``snap-shot''
observations using the Spitzer Space Telescope \citep{shafter11a}. 
These authors found evidence for dust formation in M31N 2006-10a
and (possibly) 2007-07f. M31N 2006-10a, with a $t_2$ in
the $V$ band of 83~d, was observed $\sim116$~d after
discovery, and M31N 2007-07f, with a $t_2$ of 50~d in the
$R$ band, was observed $\sim203$~d after discovery.

By contrast, we have reported evidence for dust in
\aabo\ much earlier -- 36~d after discovery, 
and only 17~d after the suggested onset of dust formation \cite{karambelkar26}. Our study therefore constitutes the
earliest, and possibly the only, dust-forming nova in M31,
in which the dust evolution has been studied at more than
one epoch. Furthermore, with a $t_2$ of 4.64~d and dust
formation in 18~d, \aabo\ falls well within the expected
region of the dust condensation time-speed class 
parameter space, as shown in Figure~2 of \cite{williams13}.

We are led to the conclusion that \aabo\ is a
luminous dusty nova in M31. Further observations are
urged to test this conclusion.
\resp{For example, timely IR spectroscopy of the dust
(which may be possible with the JWST NIRSpec and MIRI)
would be helpful in determining the nature of the dust,
and in the detection of fine structure lines 
\citep[such as \fion{S}{iv} 10.5\mic\ and
\fion{O}{iv} 25.9\mic; see][]{helton12} to determine
abundances.}

\section*{Acknowledgments}

\resp{We thank the referee for their helpful remarks.}

The Gemini observations were made possible by the award of 
Director's Discretionary Time proposals 
GN-2026B-DD-104 and GN-2026B-DD-107. We thank the
Gemini Director Time Review Board for its generous support.

This work has made use of data from the Asteroid 
Terrestrial-impact Last Alert System (ATLAS) project. The ATLAS
project is primarily funded through NASA grants
NN12AR55G, 80NSSC18K0284, and 80NSSC18K1575. It was partially funded
by Kepler/K2 grant J1944/80NSSC19K0112 and HST GO-15889, and STFC
grants ST/T000198/1 and ST/S006109/1.

DPKB is supported  by the Physical Research Laboratory,
Ahmedabad.
TRG is supported by the International Gemini Observatory,
a program of NSF NOIRLab, which is managed by the Association of
Universities for Research in Astronomy (AURA) under a cooperative
agreement with the U.S.  National Science Foundation, on
behalf of the Gemini partnership of Argentina, Brazil, Canada,
Chile, the Republic of Korea and the United States of America.
SS acknowledges partial support from a NASA Emerging Worlds
grant to ASU (80NSSC22K0361) as well as support to SS from
his ASU Regents' Professorship. CEW and SS acknowledge support
from JWST-GO-06866.002-A.

\section*{Data availability}
The raw data in this paper
that were obtained at the Gemini Observatory
are available from the Gemini Observatory
Archive, https://archive.gemini.edu/.
The ATLAS data are available at
https://fallingstar-data.com/forcedphot/~.

\bsp

\label{lastpage}

\end{document}